\documentclass[reprint,amsmath,amssymb,
 aps, superscriptaddress]{revtex4-2}
\usepackage{amsmath, braket, graphicx, amssymb}
\usepackage{bm}
\usepackage{dcolumn}
\usepackage[english]{babel}
\usepackage[normalem]{ulem}
\usepackage{xcolor}

\begin{document}
\date{\today}
\title{Efficient Quantum State Sample Tomography with Basis-dependent Neural-networks}
\author{Alistair W. R. Smith}
\affiliation{QOLS, Blackett Laboratory, Imperial College London SW7 2AZ, United Kingdom}
\author{Johnnie Gray}
\affiliation{Division of Chemistry and Chemical Engineering, California Institute of Technology, Pasadena, CA 91125, USA}
\affiliation{QOLS, Blackett Laboratory, Imperial College London SW7 2AZ, United Kingdom}
\author{M. S. Kim}
\affiliation{QOLS, Blackett Laboratory, Imperial College London SW7 2AZ, United Kingdom}
\begin{abstract}
We use a meta-learning neural-network approach to analyse data from a measured quantum state. Once our neural network has been trained it can be used to efficiently sample measurements of the state in measurement bases not contained in the training data. These samples can be used calculate expectation values and other useful quantities. We refer to this process as ``state sample tomography". We encode the state's measurement outcome distributions using an efficiently parametrised generative neural network. This allows each stage in the tomography process to be performed efficiently even for large systems. Our scheme is demonstrated on recent IBM Quantum devices, producing a model for a 6-qubit state's measurement outcomes with a predictive accuracy (classical fidelity) $>95\%$ for all test cases using only 100 random measurement settings as opposed to the 729 settings required for standard full tomography using local measurements. This reduction in the required number of measurements scales favourably, with training data in 200 measurement settings yielding a predictive accuracy $>92\%$ for a 10 qubit state where 59,049 settings are typically required for full local measurement-based quantum state tomography. A reduction in number of measurements by a factor, in this case, of almost 600 could allow for estimations of expectation values and state fidelities in practicable times on current quantum devices.
\end{abstract}
\maketitle

\section{Introduction}\label{sec:intro}
There are large practical hurdles to overcome when attempting to perform Quantum state tomography (QST) on noisy quantum devices. The extremely rapid scaling in the number of measurements that are required to completely specify a state, even at relatively small system sizes, makes full QST infeasible \cite{Titchener2018}. Full, direct QST on $n$ qubits requires the estimation of $4^n-1$ linearly independent Stokes parameters to specify a completely general quantum state and is typically performed using local projective measurements onto the eigenstates of all $3^n$ non-identity-containing Pauli strings \cite{Altepeter2005}. While this set of measurements is over-complete (containing $6^n$ quantities in total), this informationally complete protocol allows the state to be specified with the smallest number of measurement settings possible when using local measurements and so is the most effective way of implementing full tomography on a general quantum state \cite{Altepeter2005}. More efficient protocols exist \cite{Appleby2007} that do not involve collection over-complete sets of measurement data however these invariably require measurements onto entangled states \cite{Wiesniak2011}. While these entangled measurements can be indirectly performed on-device through the application of a non-local unitary before the local measurement \cite{Yen2020}, the imperfect entangling gates that are required introduce significant amounts of noise. As this noise changes the on-device state, measurements taken in different entangled bases will not properly correspond to the original desired state and so these measurement-efficient schemes are not typically implemented on current devices.
These measurements must each be repeated on many copies of the state to build up sufficient statistics to accurately estimate the Stokes parameters. This is usually followed by a maximum likelihood estimation (MLE) procedure \cite{Altepeter2005, James2001, Hradil1997, Ferrie2018} to ensure the quantum state that is inferred is valid. As the MLE is constructing a density matrix whose dimension scales exponentially with the system size, this step quickly becomes prohibitively expensive.\par
The noise on present-day quantum devices inevitably leads to errors in running the quantum circuits that produce the states. Even if we knew that an ideal preparation of the state could be reconstructed with a more limited set of measurement bases, the complexity and unpredictability of these noise processes make it difficult to know \textit{a priori} which limited measurement set would be most informative. To illustrate the scale of this issue, we can consider the time cost of performing full QST on a 10 qubit state. In our experiments we found that a single circuit on an IBM Quantum device takes on the order of 1ms to calibrate and run. Running each of the required $3^{10}=59,049$ circuits 8,192 times (the maximum allowed shot count per circuit) to build up reasonably accurate measurement statistics it would take around 130 hours to collect the data needed for full QST. These devices are frequently re-calibrated and their errors drift over time meaning that the states measured at the end of this process may vary significantly from those prepared at the start \cite{Tannu2019}. On devices that are in high demand long queue times can compound this problem even further, making full QST completely impractical for systems of this size.\par
Many physically relevant states have some underlying structure (often due to symmetries in the system) meaning that an $n$-qubit state can be described with fewer parameters than the worst-case $4^n-1$ \cite{Carleo2017, Deng2017, Bridgeman2017, Verstraete2006}. Numerous tomographic methods have been proposed that exploit this structure in order to perform tomography more efficiently, describing the state with fewer parameters which in turn require fewer measurements to learn \cite{Torlai2018, Cramer2010, Lanyon2017, Carrasquilla2019}. These concise descriptions typically provide more efficient ways of manipulating the description of the state than is possible with an explicit density matrix. Recent examples of these include machine-learning (ML) inspired approaches that attempt to take advantage of the ability of neural networks to efficiently represent and learn complicated probability distributions. These approaches have been shown to perform quantum state tomography very effectively for systems of small to intermediate size, requiring fewer measurements and being more computationally efficient than standard MLE-based techniques \cite{ahmed2020quantum}. An early example of such a scheme was proposed by Torlai \cite{Torlai2018} and is based on the ``neural-network quantum state" representation introduced by Carleo and Troyer \cite{Carleo2017}. It involves representing a pure quantum state in terms of two Restricted Boltzmann Machines (RBM, a popular generative neural network architecture). This scheme has many useful properties; for many states it can provide a very concise representation of the state (although to calculate the state exactly an exponentially costly partition function must be found), it can represent states possessing long-range entanglement, the optimisation of the ansatz is very computationally efficient for suitable states (using Hinton's contrastive divergence algorithm \cite{Hinton2002}), and as a generative neural-network it is simple and computationally efficient to draw samples from the state in the computational basis \cite{Gao2017, Deng2017-2, Torlai2018}. By efficient what is meant here is that the computational cost of each step in these processes scale polynomially with the number of subsystems/qubits.\par
Neural network (NN) approaches that attempt to produce an explicit description of quantum states in terms of density matrices or pure state vectors run into difficulties for two reasons. First, density matrices are unwieldy objects, their size scaling exponentially in the number of subsystems. This means that using them to calculate quantities, either for optimising the ansatz or analysing the inferred state, involves sums of exponentially many (potentially interfering) terms. If one is able to draw samples from the represented state in the relevant bases then these quantities can be efficiently calculated and this is the route that neural network approaches tend to take \cite{Torlai2018, Torlai2018-2, Carrasquilla2019}. However, this then leads to the second issue; density matrix and state vector descriptions of quantum states involve complex valued components whereas neural networks are typically designed to encode real valued distributions \cite{Zimmermann2011}. These complex values are required to determine measurement distributions in different bases. If one tries to account for this by using neural networks with complex valued parameters (as in with the neural-network quantum state representation given in \cite{Carleo2017}) or treating the moduli and phase of state components separately \cite{Torlai2018}, then it becomes difficult to train the network efficiently and draw samples in bases that differ greatly from the computational basis (the cost of doing so scales exponentially with the number of qubits). \par
\begin{figure*}
\includegraphics[scale=0.5]{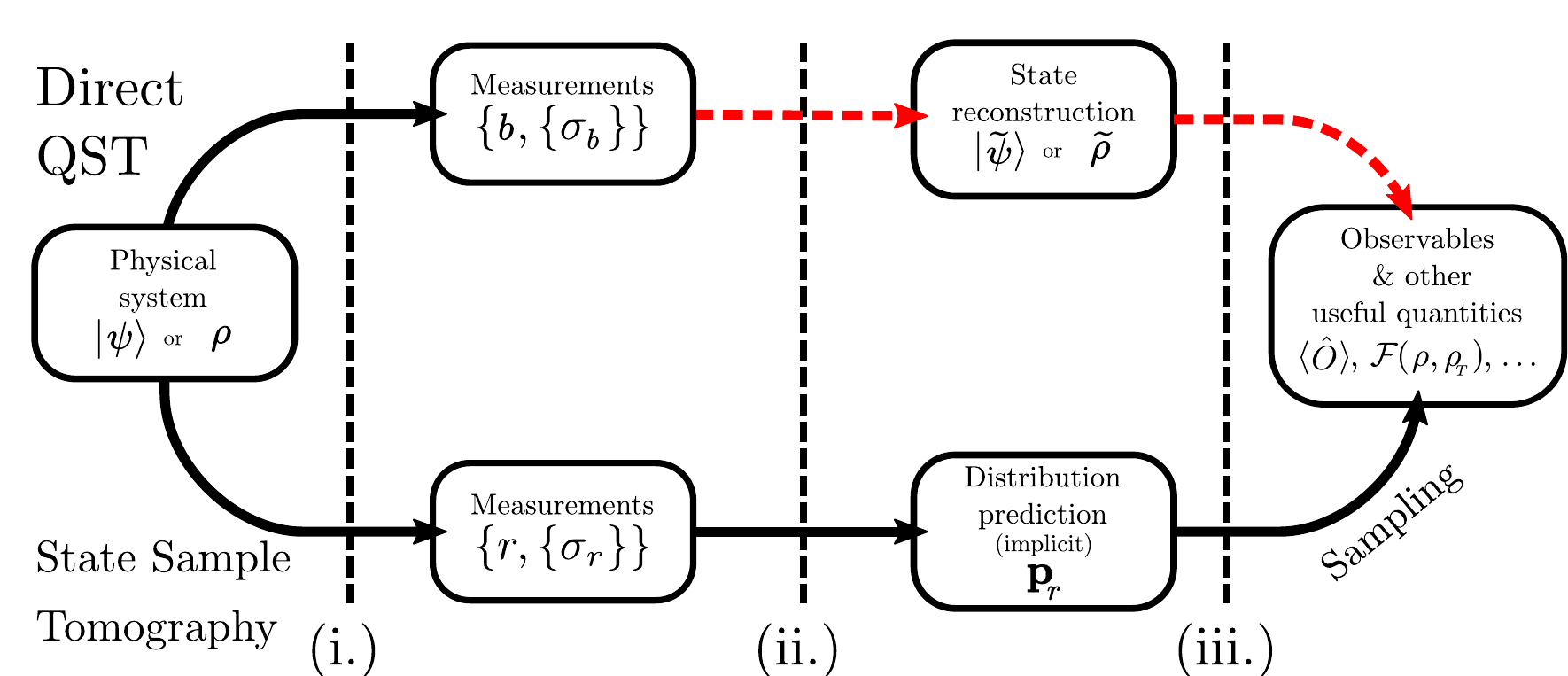}
\caption{\label{process}\textbf{Illustration of direct QST vs. SST}: Here we show the differences between direct QST, in which the density matrix is inferred, and state sample tomography where we build a model to implicitly predict measurement outcome distributions and then sample from them efficiently. The end goal in both cases is the calculation of quantities such as observables $\braket{\hat{O}}$ and state fidelities $\mathcal{F(\rho,\rho_\textit{T})}$. Lines in red show stages that become intractable for large systems when using direct QST. \textbf{(i.)} Measurements $\{ \sigma_b\}$ ($\{ \sigma_r\}$) are taken of the quantum state in various bases $\{ b\}$ ($\{ r\}$) -- the choice of bases depends on the approach taken. Whether direct QST or SST is inherently more expensive in the required number of measurements is beyond the scope of this work but in general depends on the exact methods used and the state in question. \textbf{(ii.)} Reconstruction/training stage: In SST a model is trained to map measurement settings $r$ to outcome distributions $\mathbf{p}_r$; the BDRBM model that we present performs this stage efficiently provided each local measurement distribution can be tractably approximated by an RBM. Direct QST in this stage involves optimisation over an exponentially large number of, potentially complex-valued, parameters to produce an exponentially large density matrix. \textbf{(iii.)} Calculation stage: Once the tomography has produced a model of either the state or its measurement distributions, this is used to calculate some desired quantities, which in the setting presented here are those that can be calculated efficiently from measurement results in local bases. SST allows these to be efficiently calculated by sampling from the predicted measurement distributions. Direct QST requires manipulation of large density matrices that are difficult to manipulate and sample from.}
\end{figure*}
Rather than trying to reconcile this fundamental difference between neural networks and quantum states we attempt to sidestep the issue, along with the scaling problems that can make explicit density matrices impractical to use. We skip the density matrix reconstruction stage and instead build a model of how the measurement outcome distributions vary with the choice of measurement basis. The model can then make predictions of the measurement outcome distributions in arbitrary local basis (we refer to this as ``distribution prediction"). Calculations using the full distribution of outcome probabilities quickly become intractable for large systems (there are $2^n$ outcome probabilities for $n$ qubits), making the full explicit outcome distribution of limited utility. For many parameterised probability distributions a regularisation step (calculation of a partition function) is needed to obtain the exact probabilities -- for large systems this step becomes intractable. To avoid the difficulties in using full explicit outcome distributions we use a model in which this distribution is implicitly defined in terms of an efficient number of parameters; the unnormalised probabilities of individual outcomes can be efficiently calculated but the normalisation requires an inefficient partition function calculation. While explicit probabilities are not easily accessible, as in the approach taken by Torlai \cite{Torlai2018}, we can draw samples from the implicitly represented measurement outcome distributions in each basis efficiently, and then use these to calculate quantities of interest -- we refer to this as ``sample prediction". We call the overall process of implicitly predicting the measurement outcome distribution and then sampling ``state sample tomography" (SST, illustrated in Figure \ref{process}).\par
This idea of ``sample tomography" is similar to the ``indirect"  or ``shadow" tomography discussed by Aaronson \cite{Aaronson2007, Aaronson2018} that focuses on learning the measurement outcomes of a state rather than its explicit representation. He showed that if a ``hypothesis state" can be found that is consistent with a set of observed two-outcome positive operator valued measurements (POVMs) chosen from a (potentially unknown) distribution then almost all other measurement results from this distribution can be accurately predicted with the size of this observed set scaling only linearly in the number of qubits \cite{Aaronson2007}. However, finding this hypothesis state is still computationally problematic \cite{Aaronson2007} and so we take things a step further; doing away with the hypothesis state and predicting the measurement results directly through other means (using a NN). Given that on current devices we can perform $2^n$-outcome measurements in local product bases, we focus on building a generative model for $n$-qubit measurement distributions rather than predicting two-outcome POVMs. The overall philosophy is the same; if we can \emph{implicitly} predict measurement outcome distributions (such that this process remains tractable for large systems) in arbitrary local bases and efficiently draw samples from these distributions then we have achieved a useful form of tomography.\par

Many quantum states can be verified without full QST. For example, a known target state $\rho_T$ can be decomposed onto a local operator basis (e.g. the Pauli basis $\rho_T=\sum_{\{\sigma_i\}}a_i\sigma_i$) and form a of the fidelity between the target state and the state produced on-device can be obtained by performing local measurements of these operators ($\mathcal{F}=\text{Tr}(\rho\rho_T)=\sum_{\{\sigma_i\}}a_i\braket{\sigma_i}$). Depending on the target state this may require measurements in far fewer bases than are needed for full QST. The performance of quantum algorithms is also often be quantified without full QST -- the huge measurement and classical optimisation cost of which makes it impractical for benchmarking on multiple noisy qubits. A highly relevant class of quantum algorithms to NISQ computing are variational quantum algorithms, in which a parameterised ansatz circuit is optimized to prepare a state that minimizes an efficiently measureable (on a quantum computer) cost function. As these algorithms aim to eventually solve problems that require a classically intractable number of qubits, the prepared solution state of these algorithms is rarely intended to be found exactly. Instead samples of the state in the computational basis (as in QAOA \cite{Farhi2014}), the minimum value of the cost function (as in VQE \cite{Kandala2017}), or a circuit that produces the cost-minimizing state are the intended results of the algorithms. Having a circuit to prepare the solution state still allows it to be used in further simulations/algorithms or analysed (even if full tomography is intractable).

Keeping in mind all these practical limitations, we build a generative model for local measurement distributions using what we call a BDRBM (Basis Dependent RBM). This is a RBM neural network with parameters that are basis-dependent, allowing the RBM's underlying probability distribution to continuously vary with the basis choice. A feed-forward neural network (FFNN, sometimes also called a multi-layer perceptron) is used to predict the RBM parameters for an input basis setting, these parameters implicitly determine the probability distribution represented by the RBM and so in doing this we perform distribution prediction. Sampling prediction for the measurement basis is achieved through computationally efficient Gibbs sampling of the predicted distribution. By training the two networks using measurements of the quantum state in random local bases we can do sample tomography of the state efficiently. The relationship between the two networks is illustrated in terms of the BDRBM training process in Figure \ref{bdrbm_diagram}. The FFNN can interpolate between measurement basis settings it has been trained on. This means that as long as the function mapping measurement basis choice to the corresponding RBM parameters is sufficiently smooth, we can train on a small (compared to full tomography) number of random measurement settings to get a useful, sampling-based description of the state. Unlike density matrix MLE, due to the choices of network and training procedure, all stages of this process can be performed efficiently. The computational complexity of each training step for the RBM scales linearly with the number of qubits and the FFNN can be trained using efficient back-propagation. Having one neural network determine the parameters of another is a novel approach, and one that is particularly well suited to quantum problems where we are trying to model complicated but smoothly varying probability distributions.\par 
The BDRBM scheme broadly falls into the category of meta-learning; where one attempts to solve a larger problem by looking at how different instances are learned \cite{Finn2017}. In this case the FFNN learns how the RBM represents the measurement data in different local bases and tries to link them together. The approach exploits the efficiency in training and sampling from RBMs (as in Torlai's approach) while learning measurement distributions, not the state directly, to avoid needing optimization over complex parameters and reducing the required number of measurement settings. This then allows samples to be taken efficiently in arbitrary local bases (not just the computational) giving access to all the quantities that can be calculated from these while avoiding sums over exponentially many terms. Important such quantities are is the expectation values of observables, these can be decomposed onto a set of local operators whose expectation values can be estimated efficiently through sampling in their eigenbasis. Cross-correlations of samples in random local bases \cite{Elben2020} can be used to calculate other quantities such as a form of quantum state fidelity as well as the purity of the state. As full tomography can be performed using local measurements taken in an informationally complete set of bases \cite{Altepeter2005}, by sampling from the BDRBM in this set of bases one could use it to perform full QST without having to make additional queries to the device (provided the BDRBM has learned enough about the state). As we only attempt to predict measurement outcomes, our method works in exactly the same manner for both pure and mixed states, the purity being reflected in the distributions used for training (and can be recovered with the sampling method in \cite{Elben2020}).\par

\section{Results}
\subsection{BDRBM Sample Tomography}\label{subsec:theory}
A BDRBM is a composition of two neural networks, an RBM and an FFNN. The FFNN converts measurement basis settings to a set of predicted parameters for the RBM. These parameters determine the probability distribution that the RBM succinctly represents, in this case an estimate of what the measurement distribution would look like in the local basis of interest. Once the RBM is passed the parameters predicted by the FFNN, the resultant distribution can be efficiently sampled from in an arbitrary local basis.\par
For a target state $\rho$ (as viewed in the computational basis), we build our model to predict the state's measurement distributions in different local bases. The measurement outcomes for each qubit $i$ are the $\pm1$ eigenstates $\ket{0^{(i)}_b}$ and $\ket{1^{(i)}_b}$ of a single-qubit operator $\hat{O}^{(i)}_b$ allowing the basis for that qubit to be defined in terms of the Bloch-sphere angles of the state $\ket{0^{(i)}_b}$, $\theta_i, \phi_i$. The overall basis is then specified by the set of these angles for each qubit $b=(\theta_0, \phi_0, \dots, \theta_{n-1},\phi_{n-1})$. We write the vector of $n$-qubit measurement outcome probabilities as $\mathbf{p}_b$; the $j$th component of this is the probability of the outcome $\ket{j_b}$ where $j$ is expressed in binary (e.g. for three qubits, $(\mathbf{p}_b)_{110}$ is the probability of measuring $\ket{1_b^{(2)}}\otimes\ket{1_b^{(1)}}\otimes\ket{0_b^{(0)}}=\ket{110_b}$). By rotating the state with a unitary $U(b)$ such that the basis $b$ aligns with the computational axis, $\mathbf{p}_b$ is given by the diagonal of the resulting density matrix;
\begin{equation}
\mathbf{p}_b=\text{diag}\!\left( U(b)\ \rho\ U^\dag(b) \right), \label{eq:mapfn}
\end{equation}
\begin{equation}
U(b)=\bigotimes_{i=0}^{n-1} U_i(\theta_i, \phi_i),
\end{equation}
\begin{equation}
U_i(\theta_i, \phi_i)=\begin{pmatrix}
\cos{(\frac{\theta_i}{2})} & e^{-i\phi_i}\sin{(\frac{\theta_i}{2})}\\
\sin{(\frac{\theta_i}{2})} & -e^{-i\phi_i}\cos{(\frac{\theta_i}{2})}
\end{pmatrix}.\label{eq:rots}
\end{equation}

The $2^n$ functions $\mathbf{p}_b$ are generally complicated as the unitaries $U(b)$ allow a potentially large number of off-diagonal terms to contribute. The analytic form of these equations is cumbersome and entirely dependent on the state in question, making them difficult to exploit in a general tomographic scheme, particularly when dealing with an unknown state. Instead we note that they are continuous functions of $\{\theta_i, \phi_i\}$ and take advantage of the flexibility of neural networks as function approximators to try to infer them from measurement data. As we have moved away from the density matrix formalism, the approach comes with the caveat that while the BDRBM gives a valid probability distribution in any local basis, the distributions in different bases do not necessarily reconcile to form valid quantum state. However, provided the BDRBM sufficiently expressive and is trained on enough data that it yields good accuracy in predicting the measurement outcome distributions for the state then the state's properties should be (at least approximately) reproduced. The BDRBM approach assumes that the measurement distribution of the state in question changes relatively smoothly as the basis is varied allowing interpolation between the settings used for training. We justify by noting that the rotation matrices in Eq.(\ref{eq:rots}) are smooth functions of the measurement basis angles $\{\theta_i,\phi_i\}$ and so measurements in a given basis provide information about other nearby bases. A basic requirement for this is that the training data covers each qubits' Bloch sphere sufficiently. The RBM can identify correlations in measurement data but these correlations vary with the basis setting. This variation must be learned by the FFNN and so enough training data is required that this can be done.\par
The measurement outcome distribution (for a given basis) in our BDRBM model is represented using an RBM. RBMs were first introduced by Smolensky \cite{Smolensky1986}, and have become a staple tool in the ML community following Hinton's invention of the contrastive divergence training algorithm \cite{Hinton2002}. They consist of an undirected graph of two connected layers of binary valued ``neurons" or ``nodes"; a ``visible" layer of size $n_{\bf{v}}$ representing the states of the physical subsystems (in this case the qubit states after measurement in the basis of interest) and a stochastic ``hidden" layer (of size $n_{\bf{h}}$). Each configuration of neuron states has an associated energy $E(\bf{v},\bf{h})$ and a probability of occurring given by $p(\mathbf{v}, \mathbf{h})=e^{-(E(\bf{v}, \bf{h}))}$. Assuming that the distribution in question has some exploitable underlying structure, RBMs are a parameter-efficient way to encode a probability distribution compressing $2^{n_v}$ outcome probabilities into $n_v+n_h+n_vn_h$ RBM parameters. The probability of a given configuration of visible neuron states $\bf{v}$ occurring is determined by the values of a set of learned RBM parameters $\lambda$; vectors $\bf{b}$ and  $\bf{c}$ (bias terms on the visible and hidden neurons), and a weight matrix $\bf{W}$ that mediates correlations in the distribution. The unnormalised probability $\tilde{p}(\mathbf{v})$ of a given configuration of visible neurons, $\mathbf{v}$ is given by
\begin{equation}
\label{eq:rbmProb}
\tilde{p}(\mathbf{v})=e^{\sum_ib_iv_i}\prod_j\left(1+e^{c_j+\sum_iv_iW_{ij}}\right).
\end{equation}
The actual probability of this configuration occurring is then obtained by normalising this quantity;
\begin{equation}
\label{eq:rbmnorm}
\begin{split}
p(\mathbf{v})&=\frac{1}{Z}\tilde{p}(\mathbf{v})\\
Z&=\sum_{\{\mathbf{v}\}}\tilde{p}(\mathbf{v}).
\end{split}
\end{equation}
Direct calculation of this probability requires first calculating the partition function $Z$ -- as this involves a summation over $2^{n_{\mathbf{v}}}$ possible configurations of $\mathbf{v}$ this becomes intractable for large systems (large $n_\mathbf{v}$). A computationally-efficient algorithm was proposed by Hinton \cite{Hinton2002} to optimise the RBM parameters $\bf{b}$, $\bf{c}$, and $\bf{W}$ to minimise the Kullback-Leibler divergence between the model's probability distribution and the distribution of a set of training data. This is done by making use of a Markov-chain Monte Carlo (MCMC) process that efficiently generates samples from the RBM's represented distribution while avoiding calculation of $Z$ \cite{Torlai2018}. The time taken for these samples to converge to the RBM's underlying distribution (the asymptotic distribution of the MCMC process) varies depending on the implementation of the RBM, the distribution of the starting samples, and the complexity of the underlying distribution and so we will not cover it in detail here; it is however an important topic of ongoing research \cite{Hinton2002, Breuleux2011}. Crucially, each step of the MCMC process scales only as $\mathcal{O}(n_{\mathbf{v}}n_{\mathbf{h}})$ and so the training and sampling processes can be extremely computationally efficient provided that $n_{\mathbf{h}}$ is not too large. RBMs are capable of approximating arbitrary probability distributions, provided that the target distributions are sufficiently smooth and regular \cite{Jia2019}. However, this may require an exponentially large number of hidden neurons, making the representation impractical \cite{Jia2019}. This places some practical limitations on what a BDRBM is capable of but as most states of interest in many-body physics possess numerous symmetries or underlying structure this regularity condition seems not too great a requirement.\par 
We increase the representational power of the network by allowing the RBM's parameters to change with the basis choice, $\lambda=(\mathbf{b}, \mathbf{c}, \mathbf{W})\rightarrow\lambda(\mathbf{r})=(\mathbf{b}(\mathbf{r}), \mathbf{c}(\mathbf{r}), \mathbf{W}(\mathbf{r}))$. To do this in a generalised framework we use a feed-forward neural network (FFNN). This is a simple but highly expressive neural network architecture consisting of layers of real valued neurons with (forward-flowing) connections between them. The value of neurons in each layer are calculated by acting with an (usually non-linear) activation function on the weighted sum of the neuron states in the previous layer, allowing an input state to be converted to an output state. In this case the FFNN takes as its input desired measurement basis and outputs a set of RBM parameters. We smoothly parametrise the basis on each qubit using the Bloch sphere coordinates of the normalised measurement axis vector, $\textbf{r}_i=(x_i,y_i,z_i)$; a more concise description would instead use the polar angles of this axis however Cartesian coordinates were found to work more effectively (while also avoiding discontinuities). The output of an $m$-layer FFNN with $(n_{l_1}, \dots, n_{l_m})$ hidden neurons in each layer, $\lambda(\mathbf{r})$, is then given by
\begin{equation}
\lambda(\mathbf{r})_i=\lambda^{out}_i + K^{out}_{ij}\left((g_m\circ g_{m-1}\circ \dots \circ g_1)( \mathbf{r})\right)_j,
\end{equation}
\begin{equation}
g_k: \mathbb{R}^{n_{l_{k-1}}} \rightarrow \mathbb{R}^{n_{l_k}}; \mathbf{s} \mapsto f_k(\mathbf{a}_k + \mathbf{K}_k\cdot\mathbf{s}),
\end{equation}
i.e. a composition ($\circ$ is the composition operator) of functions $\{g_k\}$ that map the input of one layer to the output of the next. Overall, this maps the input $\mathbf{r}$ to an output $\lambda(\mathbf{r})$. Each $g_k$ involves an element-wise activation function $f_k$, a bias vector $\mathbf{a}_k$, and a weight matrix $\mathbf{K}_k$. To allow the FFNN to act as a universal function approximator these activation functions should in general be non-linear. In experiments found that while a linear FFNN was effective in many cases, the popular $leaky\_relu$ function, $g(x)=\text{max}(x, 0.2x)$ \cite{Maas2013} also performed well (allowing for some degree of non-linearity).\par
The exact form of the FFNN for a given state, i.e. number and size of the hidden layers and their activation functions, should be chosen on a case-by-case basis. As previously discussed, while the networks produce valid probability distributions in arbitrary local bases, these will be an approximation to the statistics of the density matrix and may not exactly correspond to a positive-definite quantum state. On the other hand, the distributions in bases that are different from the computational basis are more directly accessible than when using a density matrix as the RBM parameters that define this can be obtained with a single pass through the FFNN, rather than a computationally expensive rotation of the state. This means that quantities that can be calculated using samples in local bases (e.g. state entropy, certain state fidelities, or physical observables) may be estimated more tractably with a BDRBM than would be feasible with exact calculations on a large density matrix.\par
\begin{figure*}
\includegraphics[scale=0.6]{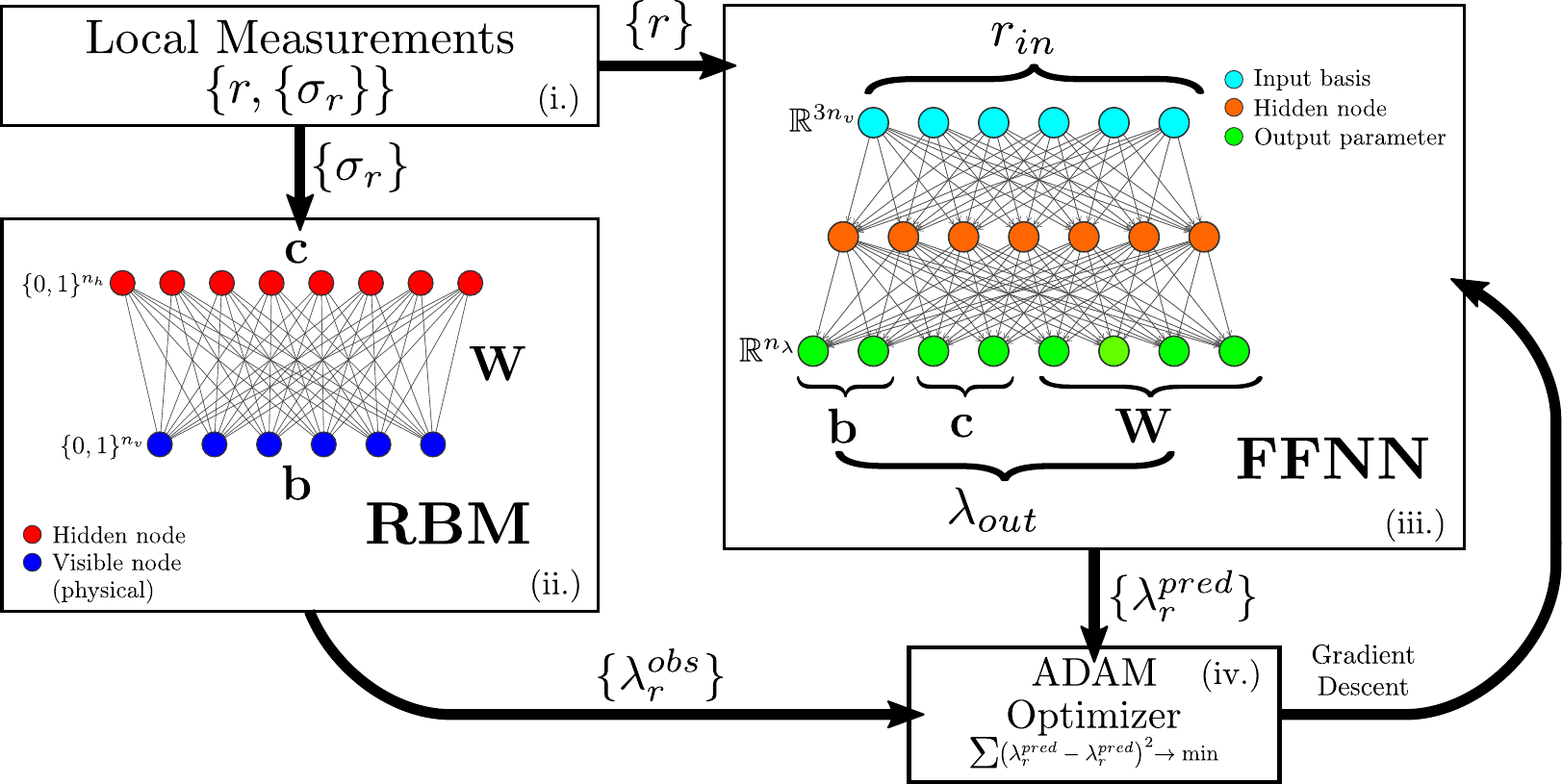}
\caption{\label{bdrbm_diagram}\textbf{Diagram of BDRBM training process.} \textbf{(i.)} Measurements $\{\sigma_r\}$ are taken of the quantum system in local bases denoted by $\{r\}$, the Bloch sphere coordinates of the each qubits' measurement axis. The measurement bases are randomly chosen such that $\{r\}$ is uniformly distributed on the surface of the Bloch (upper hemi)sphere for each qubit. \textbf{(ii.)} These measurement results are then used to train an RBM with the CD-1 algorithm, the RBM parameters are recorded as $\lambda_r^{obs}$.  \textbf{(iii.)} A FFNN (here shown with one hidden layer) takes the basis setting as an input and outputs a set of predicted RBM parameters $\lambda_r^{pred}$, thereby implicitly predicting the measurement distribution for this basis setting. \textbf{(iv.)} The parameters in the FFNN are trained using gradient descent, minimising the squared difference between the $\lambda_r^{pred}$ and $\lambda_r^{obs}$.}
\end{figure*}
Training the network takes place in several stages, as illustrated in Figure \ref{bdrbm_diagram}. First, measurements are performed on the quantum system in a number of local bases; to ensure that the FFNN can predict across the desired space of basis settings these bases must reasonably well cover the Bloch spheres of each qubit. We observed that the network predicted distributions most accurately when the bases were selected at random uniformly across each qubit's Bloch sphere. Points on opposite sides of the spheres give equivalent measurements (up to relabelling) meaning that the measurements can be restricted to only the Bloch upper-hemispheres, reducing the amount of data needed to learn $\lambda{(r)}$. The basis settings are retained to form the inputs for the FFNN's training data. For each measurement the RBM is trained to find a compressed representation of the measurement data (using Hinton's CD-1 algorithm \cite{Hinton2002}) and the observed RBM weights and biases for each basis are recorded to form the outputs of the FFNN's training data $\{\lambda^{obs}_r\}$. To ensure similarity and smoothness between observed RBM parameters from basis-to-basis, and to speed up convergence, the parameters found for one basis are used as the starting point for the next. L2-regularisation (adding a cost penalty to the summed square weight elements) is used during this stage to prevent the RBM weights growing too large; when done correctly this both speeds up and improves convergence \cite{Hinton2012} while leading to more manageable variation within the FFNN's training data.\par
Finally, regression is performed using gradient descent (the ADAM algorithm was used in the examples given \cite{ADAM}) on the observed RBM parameters to fit the FFNN, minimising the $\ell^2$-norm between observed and predicted parameters (along with some L1-regularisation of the FFNN weights to reduce over-fitting and allow clearer analysis of the weights). An iterative fine-tuning process was found to reduce over-fitting and lead to better accuracy in the predicted distributions, this is done by retraining the RBM on the measurement data using the FFNN's predictions for each training basis as the RBM's starting point. RBMs are very flexible models and so can find many different but similarly faithful representations of a distribution. It is likely that during the initial RBM optimisation step parameters will be observed that may not reconcile well with the rest of the data set. Iteratively predicting and re-learning the RBM parameters allows the observed parameters to drift closer to a set that can be most effectively described by the FFNN.\par
It is often the relative sizes of the RBM parameters that are most important in representing a given distribution and so the parameters often exhibit a large amount of covariance. This can be exploited in an optional pre-processing step prior to the regression using principal component analysis (PCA). PCA finds the linear combinations of RBM parameters that contribute the most to the covariance of the data and ranks them in terms of this contribution \cite{Lever2017}. By rewriting the RBM parameters in terms of their projections onto these directions the regression process can converge more quickly and accurately. In addition, by only retaining the projections onto the directions that explain the largest amounts of covariance one can greatly reduce the dimension of the training data, potentially without losing much useful information \cite{Lever2017}. This results in a smaller FFNN being required making both the regression step easier and leading to a more concise final representation of the state. \par
\subsection{Classical Fidelity for Noisy Quantum Device TFIM Generation}\label{subsec:fid}
To demonstrate this tomographic process on a real quantum system we applied it to random local basis measurements taken for a set of 6-qubit states prepared on the $\textit{ibmq\_singapore}$ and $\textit{ibmq\_paris}$ devices \cite{Qiskit} (run in early April 2020). These states were prepared using circuits that were optimised with a tensor-network based method (using the quimb python package \cite{Gray2018}) to output a state that, in the absence of any device errors, matches the ground-state of the 1-D anti-ferromagnetic transverse field Ising model (TFIM), to within a quantum infidelity of order $10^{-5}$. This was done by expressing an ansatz circuit in terms of a tensor-network and then taking advantage of tensor contraction schemes to maximise the fidelity between the target ground state and the output of the circuit. The anti-ferromagnetic TFIM Hamiltonian used here is for a $1$-dimensional spin-chain with spins on the $i$th site $S_i^x$ and $S_i^z$ along the $x$ and $z$ axes respectively is given by
\begin{equation}\label{eq:tfim}
H=J_z\sum_{\langle i,j\rangle}S^z_iS^z_j-J_x\sum_iS^x_i,
\end{equation}
where the first sum is performed over nearest neighbour pairs $\langle i,j\rangle$. The system undergoes a phase transition when $J_z$ (the interaction strength) and $J_x$ (the transverse field) are equal in magnitude. This model provides a good test case for the tomographic process as by varying the relative size of $J_z$ and $J_x$ we can examine its performance as the ground-state goes from the Bell-like state ($\ket{\psi_0}\sim\ket{01}^{\otimes n/2}+\ket{10}^{\otimes n/2}$) for small $J_x$ to the non-entangled product state ($\ket{\psi_0}\sim\ket{+}^{\otimes n}$) via an intermediate state that exhibits long-range entanglement. The BDRBM utilised here consisted of an RBM with 6 hidden neurons and a simple linear FFNN (consisting only of input and output layers) which outputs RBM parameters, $\lambda^{out}$, as
\begin{equation}
\lambda^{out}_i(\mathbf{r})=\lambda^0_i+\sum_{j}M_{ij}r_j.
\end{equation}\par
Training the BDRBM to recreate measurement outcome distributions for data in bases that it has already seen is not enough for useful tomography. It is also necessary that the predicted distributions for unseen bases are accurate and so we need to be able to test how well this is done. If we were using a density matrix ansatz then we could simply calculate the fidelity between the learned and target state; however, as discussed in \ref{sec:intro}, the size of the density matrices makes this impractical. To check that the network's performance on unseen data is similar to that on the training data we use a simple cross-validation method. This involves retaining a portion of the measurement data (i.e. not using the data for these bases to train the BDRBM) and comparing the observed probabilities of each outcome to the predictions for the held-out measurement bases. There are several possible figures of merit for the comparison between the predicted and observed distributions. The one that is used here is the classical version of the quantum fidelity of a pure state (the quantum fidelity can be shown to be the maximisation of this quantity across all possible POVMs). The ``classical fidelity", $\mathcal{F}_c$, (called the Bhattacharyya coefficient in classical statistics) expresses the similarity between two probability vectors $\mathbf{p}$ and $\mathbf{q}$ is given by
\begin{equation}
\mathcal{F}_c(\mathbf{p},\mathbf{q}) = \sum_i\sqrt{p_iq_i}.
\end{equation}\par
The quantum fidelity can be can be estimated with local measurement samples taken from the BDRBM using the method described in \cite{Elben2020} and so could be used for cross-validation. However, for this purpose we use the classical fidelity as it is significantly faster to calculate. It also allows us to make comparisons between the BDRBM's predictions and the measured device data on a basis-by-basis level. This makes it well suited to the cross-validation used to identify if over-fitting has occurred as it gives us two well defined standalone figures of merit for the training and validation sets. The quantum fidelity is a combined comparison for all bases and so, as we are also dealing with a generative model for measurements rather than a density matrix, is not necessarily any better suited for this task. As shown in \cite{Carrasquilla2019}, while the classical fidelity provides only an upper bound on the quantum fidelity it provides an effective substitute; a high classical fidelity usually indicates a high quantum fidelity.\par 
Circuits were run on the $\textit{ibmq\_singapore}$ and $\textit{imbq\_paris}$ devices to prepare the ground states of the 6-qubit anti-ferromagnetic TFIM with varying $J_x$ (and constant $J_z=1$), as in Eq.(\ref{eq:tfim}). These states were then measured in 200 random local bases (uniformly distributed on qubit upper Bloch hemispheres). For each device, the results were split into a training and a validation set and the training data was fed into the tomographic algorithm as detailed in \ref{subsec:theory}. As we are considering only a relatively small system, we compare the exact predicted probability distributions (using Eq. \eqref{eq:rbmnorm}) to the empirical distributions in the measured data. For larger systems these comparisons could instead be made tractably (although with some introduced sampling error) between samples predicted by the BDRBM and the measurement data. To compare the performance of the algorithm for different transverse field strengths, the average classical fidelities across the bases in the training and validation sets were calculated between the observed probability distributions in the measurement data and the BDRBM's distribution predictions. As these are NISQ devices, it is also useful to see how well the BDRBM's predictions match what would be expected in the absence of device noise and so average classical fidelities were also found between the ideal measurement distributions for target state and the BDRBM's predicted distributions for the bases in the two data sets. Over-fitting by the FFNN can be quantified by looking at the extent to which the reconstructive fidelity (for the training data set) is higher than the predictive fidelity (for the unseen validation set), this is a basic form of cross-validation.\par
\begin{figure*}
\includegraphics[scale=0.6]{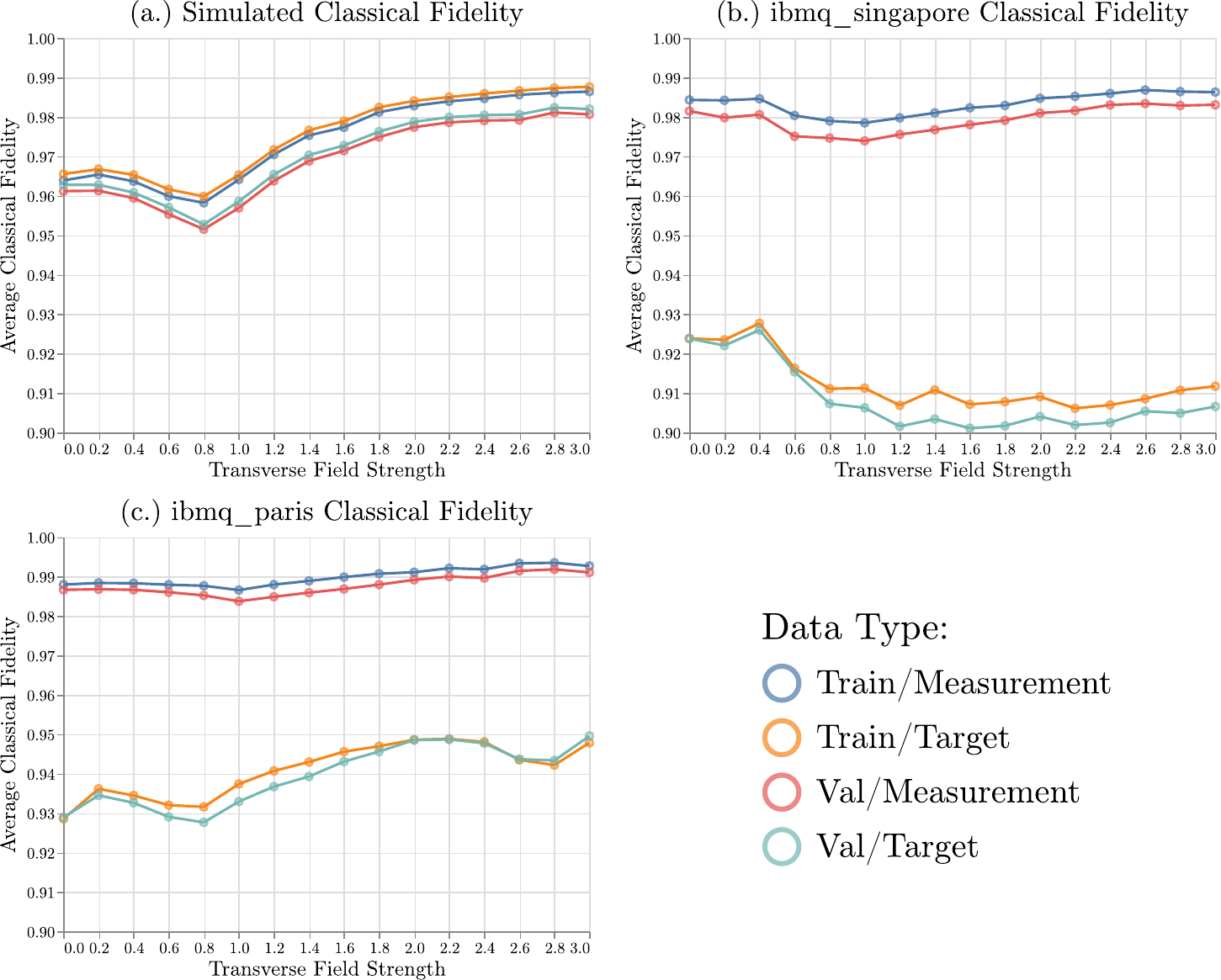}
\caption{\label{fig:fidelity_graphs}\textbf{Achieved classical fidelities for BDRBM tomography of 6-site anti-ferromagnetic TFIM.} These graphs show the average classical fidelities between the predicted measurement distributions of the BDRDBM after training with \textbf{(a.)} ideal simulated data, \textbf{(b.)} data from the $ibmq\_singapore$ device, and \textbf{(c.)} data from $ibmq\_paris$ over a range of transverse fields (and $J_z=1$). The training process was as described in section \ref{subsec:theory}, with 6 hidden neurons in the RBM and a linear FFNN with directly connected input and output layers. Each graph shows four fidelities between the BDRBM's predicted distributions and; Train/Measurement - the observed measurements in the bases used for training, Train/Target - the exact simulated target distributions for the training bases, Val/Measurement - the withheld measurements in the bases used for validation, Val/Target - the exact simulated target distributions for the validation bases.}
\end{figure*}

The achieved average classical fidelities $\overline{\mathcal{F}_c(\mathbf{p}_{obs},\mathbf{q}_{rbm})}$ (the mean of classical fidelities for the bases used in the comparison) for simulated (ideal) data and for the data collected from $\textit{ibmq\_singapore}$ and $\textit{imbq\_paris}$ are shown in Figure \ref{fig:fidelity_graphs}. These were calculated between the observed measurement distributions $\mathbf{p}_{obs}$ and the BDRBM's predicted distributions $\mathbf{p}_{rbm}$ for the bases in question as in Eq. \ref{eq:rbmnorm}.\par

Figure \ref{fig:fidelity_graphs}a shows that this simple linear BDRBM performs well for an ideal, simulated preparation of the state. Both the reconstructive and predictive fidelities are high with a small degree of over-fitting indicated by the reconstructive fidelity being slightly higher than the predictive. We see a dip in both fidelities at a transverse field strength of around $J_x=J_z=1$; at this point the ground state of the anti-ferromagnetic TFIM undergoes a quantum phase transition from an ordered to disordered phase (as viewed in the computational basis) via a state with long range entanglement. A dip in fidelity around this phase transition is expected as the more complicated entanglement structure means that in the state's Schmidt basis (the product basis with the minimum number of terms in its description of the state) a larger number of significant terms are present. All of these terms contribute to the function that maps basis choice to outcome probabilities (Eq.(\ref{eq:mapfn})), making this function more complicated therefore and harder to learn.\par
For both NISQ devices (Figures \ref{fig:fidelity_graphs}a and \ref{fig:fidelity_graphs}b) we see that the predictive fidelity is consistently high between the BDRBM's predictions and the withheld measurement data (labelled Val/Measurement in Figure \ref{fig:fidelity_graphs}) at around $97-98\%$. This is true even at transverse fields around the phase transition where the BDRBM trained with ideal simulated data only manages to achieve predictive fidelities of around $95\%$. This indicates that the true states on the devices at these points are easier to learn than the intended highly entangled ground-state, showing that the devices may not be able to reliably produce the long-range entanglement required.\par
When fed data taken on $\textit{ibmq\_singapore}$, the model reproduces both the training (reconstruction) and validation (prediction) sets' observed measurement distributions with a high classical fidelity (Figure \ref{fig:fidelity_graphs}b). However, these BDRBM-predicted measurement distributions do not agree particularly well with the measurement distributions of the ideal target state. The fidelities between these distributions are a lot lower and this does not improve as the transverse field gets large (despite the target ground-state becoming much simpler). This discrepancy indicates that the tomographic algorithm has managed to accurately learn the state of the system, incorporating the noise which leads to an imperfect preparation of the desired state. As the ground states for larger transverse fields (past the phase transition) get closer to the $\ket{+}^{\otimes n}$ state, this state consists of a superposition of all possible physical states on the device (the computational basis). In the presence of a large amount of de-phasing noise, for example due to cross-talk between qubits or large numbers of noisy 2-qubit gates, these states could prove fragile. If this noise is correlated (for example as a result of a state dependent crosstalk changing the locally experienced fields of neighbouring qubits) then the more highly entangled states (ones with low $J_x$) may be more resistant to it, leading to the higher target-prediction fidelity seen in the data for low $J_x$. The circuits used to generate these states consisted of an ansatz with a depth (the number of controlled-Z gates acting between each adjacent qubit pair) of 3; this is overly complex for the relatively simple, almost separable ground-states at large transverse fields and so excessive noise due to errors on the 2-qubit gates may have either ruined the superposition by the time the measurements take place or entangled the state more than is necessary.\par
The $\textit{ibmq\_paris}$ device appears to fare much better (Figure \ref{fig:fidelity_graphs}c) than $\textit{ibmq\_singapore}$. A discrepancy is still observed between the experimental data/prediction and target/prediction fidelities implying that a good deal of noise is still present however this discrepancy is smaller than for $\textit{ibmq\_singapore}$. Crucially the performance of the tomography increases as the transverse field becomes large, qualitatively agreeing with the trend observed in the simulated data. This indicates that the newer device is considerably better at maintaining coherences between its computational basis states. A reduction in fidelity is observed for the largest values of $J_x$, as mentioned before this is likely due to the circuit ansatz being used containing more gates than is strictly necessary for the simple states in question. As the same depth-3 ansatz was used for these circuits we again have an excessive number of 2-qubit gates that will not aid greatly in the state preparation and, as these gates have much higher errors than single qubit gates, these are likely the primary cause of the suppression of the fidelity for large $J_x$.\par
\subsection{Scaling}\label{subsec:scaling}
Analysing how many measurement settings are required for this tomographic process is difficult as there are many factors that will increase the complexity of the learning process. Broadly speaking, it will depend on the complexity of the state in question; it is expected that the number of measurements required will increase with the number of significantly contributing terms in the state's local Schmidt basis, as this makes Eq.(\ref{eq:mapfn}) more complicated. The suitability of an RBM in representing the target state in local bases is also important. While RBMs are capable of closely approximating arbitrary probability distributions this may require an infeasibly large number of hidden neurons making the learning and sampling processes too slow to be practical. The RBM must be able to well approximate the state in an arbitrary basis but it must also do so with relatively smooth changes of the parameters (which can be learned by the FFNN), this may also increase the number of hidden neurons required. This issue of parameter smoothness could be partially alleviated by using a more complicated FFNN; however, this may then increase the number of parameters to be trained and so require more training data. If dimensionality reduction is to be performed using PCA then the number of sets of training RBM parameters (and so the number of measurement settings) must be greater than or equal to the number of parameters in the RBM ($n_v+n_h+n_vn_h$) and so this puts a lower bound on the number of measurements required if this is to be used.\par
\begin{figure*}
\includegraphics[scale=0.6]{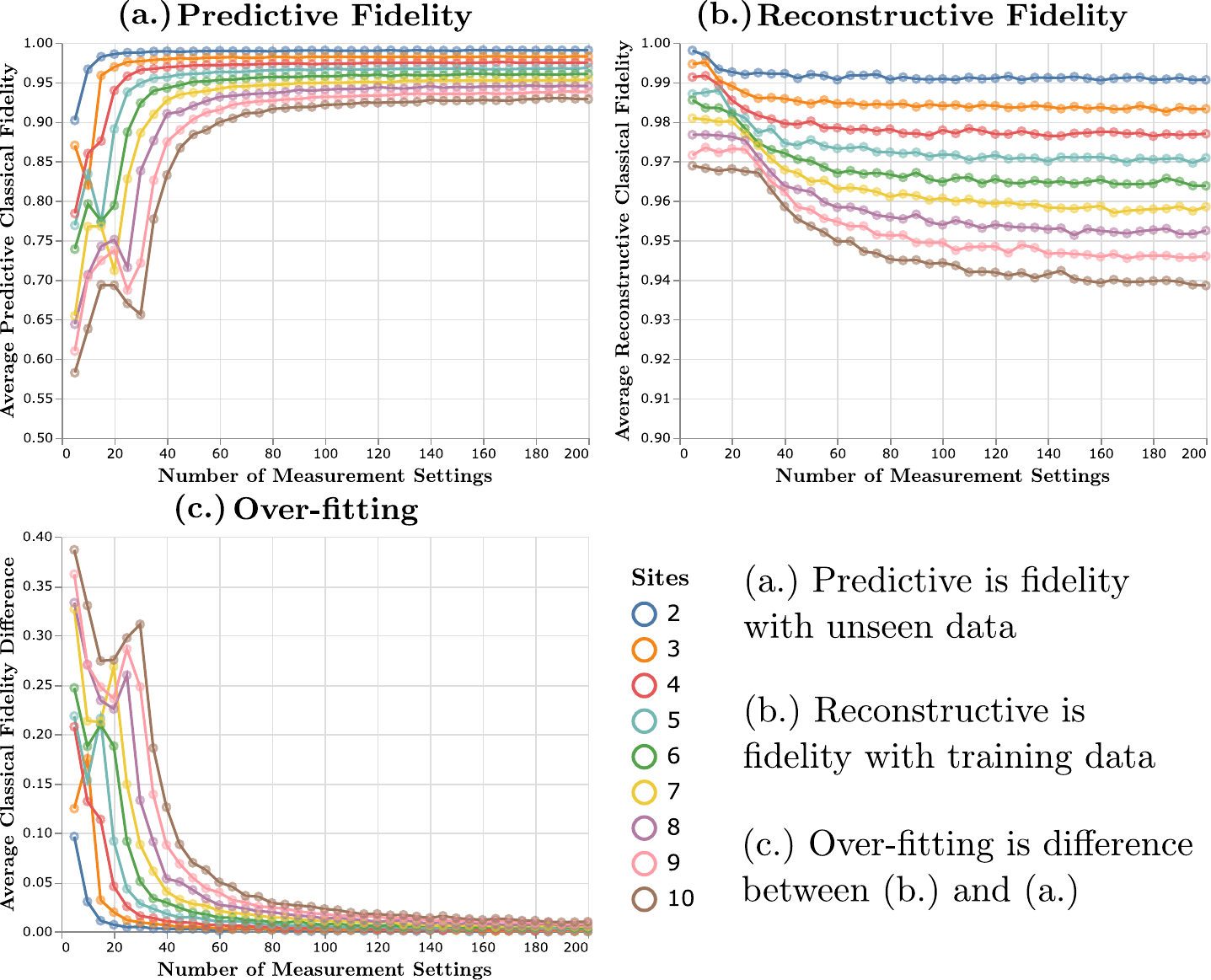}
\caption{\label{scaling_graphs}\textbf{Scaling of fidelities and over-fitting with number of training measurements.} These graphs show how performance of our scheme changes as the BDRBM is given larger sets of training data. This is shown for a linear BDRBM trained with different numbers of simulated measurements (noiseless with 8192 shots per measurement basis) of the anti-ferromagnetic TFIM ground-state with $J_x=J_z=1$ (the quantum-critical point). These simulations were performed for ground-states with a range of $2$ to $10$ sites (indicated by line colour). The RBMs used had $n_v=n_h$ \textbf{(a.)} The scaling of the average predictive classical fidelity between the BDRBM's predicted measurement distribution and the exact (calculated from the target state's state-vector) distributions in randomly selected bases that are not used for training. \textbf{(b.)} The scaling of the average reconstructive classical fidelity between the BDRBM's predicted distribution and the exact distributions in the bases used for training. \textbf{(c.)} The scaling of the difference between the average reconstructive and predictive classical fidelities, a measure of how much the network has over-fitted to the training data.}
\end{figure*}
To investigate the scaling of the required the number of measurement settings with the size of the state in question a linear BDRBM (as described in section \ref{subsec:fid}) was trained to learn the anti-ferromagnetic TFIM ground-state. The size of the state and the number of simulated measurements being passed to it were varied. The transverse field used was $J_x=1$, corresponding to the critical point of the phase transition and the region that is hardest for the network to learn. To allow the model complexity to grow suitably with the increasing size of the state the number of hidden neurons was set to be equal to the number of sites (this was found to give the best predictive fidelity). The achieved predictive and re-constructive fidelities for different system and training data sizes are shown in Figure \ref{scaling_graphs}a and \ref{scaling_graphs}b. Figure \ref{scaling_graphs}c gives the difference between the re-constructive and predictive fidelities. This indicates the amount of over-fitting that has occurred.\par
The fidelity is seen to first improve as the training set gets larger but then briefly decreases before a further rapid increase leads to saturation. Error curves of this shape (or a fidelity curve in this case) are well documented in deep-learning (and in over-parametrised regression more generally) as a phenomenon known as ``double descent" \cite{Belkin2019, Nakkiran2019}. As more training data is added the model's performance improves up to the point where the over-parametrised neural network starts to fit the training data too closely and the generalisation error increases (these are the drops in predictive fidelity in Figure \ref{scaling_graphs}a). Eventually the ``interpolation threshold" is reached and the model uses its excess parameters to interpolate directly between data points, giving the maximum amount of over-fitting (the peaks in Figure \ref{scaling_graphs}b) \cite{Nakkiran2019}. However past this point the inductive bias due to the regularisation procedure used (in our case L1-regularisation) comes into play and any added data allows the stochastic gradient descent algorithm to find the predictors that give as simple a model as possible without sacrificing too much reconstructive performance. This ``Occam's razor"-like approach then results in a greatly improved generalisation error (the rapid improvement in predictive fidelity seen in Figure \ref{scaling_graphs}a following the dips) \cite{Belkin2019}.\par
Using this linear model the predictive fidelity saturates at lower values as the number of sites grows, this implies that a more expressive model is needed to more accurately represent large ground-states of this type. However, the achieved average fidelity of around $93\%$ means that, given the size of the system, the predicted measurement distributions will exhibit the significant features of the true probability distribution, with peaks and dips in the distributions located in the correct places. This reasonably faithful approximation of the state can be achieved with only on the order of $100$ measurement bases and, in terms of the measurements taken, makes no assumptions about which terms in the density matrix (and so which measurement bases) might be relevant. In a NISQ setting where noise processes lead to substantially mixed states, full characterisation would require full tomography to be performed and this would instead require $3^{10}=59,049$ measurements. 
\subsection{TFIM Filters for Linear BDRBM}\label{subsec:tfim_filters}
When using neural networks it is often illuminating to look at the filters (the weight matrix $M$ in this case) that the model has inferred from the data to try to draw some conclusions about the underlying system. For the simple linear model used for the TFIM data this weight matrix is easy to interpret; the element $M_{ij}$ gives the derivative of the RBM parameter $i$ with respect to the basis coordinate $j$, $M_{ij}=\partial \lambda_i /\partial r_j$ and so tells us how strongly the parameter depends on that coordinate. In this section we analyse the filters found from the tomography in section \ref{subsec:fid} and discuss how these relate to the state that is learned.\par
While the non-linearity of an RBM makes it difficult to directly interpret its parameters we can still make some qualitative statements. Broadly speaking, the visible biases, $\mathbf{b}$, control the probability of their associated neurons taking the state $1$ (or $\ket{1}$ if these neurons represent a qubit state) \cite{Hinton2012}. For a product state the reduced state of a given qubit is independent of state of the others and so these could therefore be represented using an RBM with no hidden neurons and visible biases that vary only with the measurement basis of their own associated qubit. The weights and hidden biases have the role of mediating correlations between outcomes and so are needed to express more complicated states. Non-local correlations may also be indicated by visible biases (a parameter that typically defines local properties) that are dependent on the measurement settings of qubits other than the one they are associated with.\par
Following these simple qualitative interpretations, one would expect that for the anti-ferromagnetic TFIM ground states with $J_x<J_z$ the measurement distributions would be most strongly dependent on the $z$ coordinates of each qubit's basis and so the non-zero components of $M$ will mostly lie along rows corresponding to $z$ coordinates. This is because the $z$ axis corresponds to a Schmidt basis for the state in this ground-state. Measurements taken along axes close to this result in sparse distributions with sharp peaks; distributions with many zeroes and a few large peaks require the largest weights and biases for the RBM and so these axes (and therefore components along them) end up being the most important in determining the RBM parameters. In this regime the state is entangled and so we also expect to see that the columns of $M$ that correspond to hidden weights and biases contain non-zero values as these RBM parameters will be required to express correlations in the measurement outcomes (and must be able to vary with the local basis choices). The filters that we expect for large transverse fields are much simpler; as there is little entanglement in these cases we expect that the dominant terms in $M$ will correspond to visible biases and that these should only depend on the basis choice for their associated qubit. As these ground-states are close to $x$ axis eigenstates ($\ket{+}^{\otimes n}$) we would expect that the $x$ values of the basis choice to become significant, in particular determining the visible biases of their respective qubit.\par
\begin{figure*}
\includegraphics[scale=0.4]{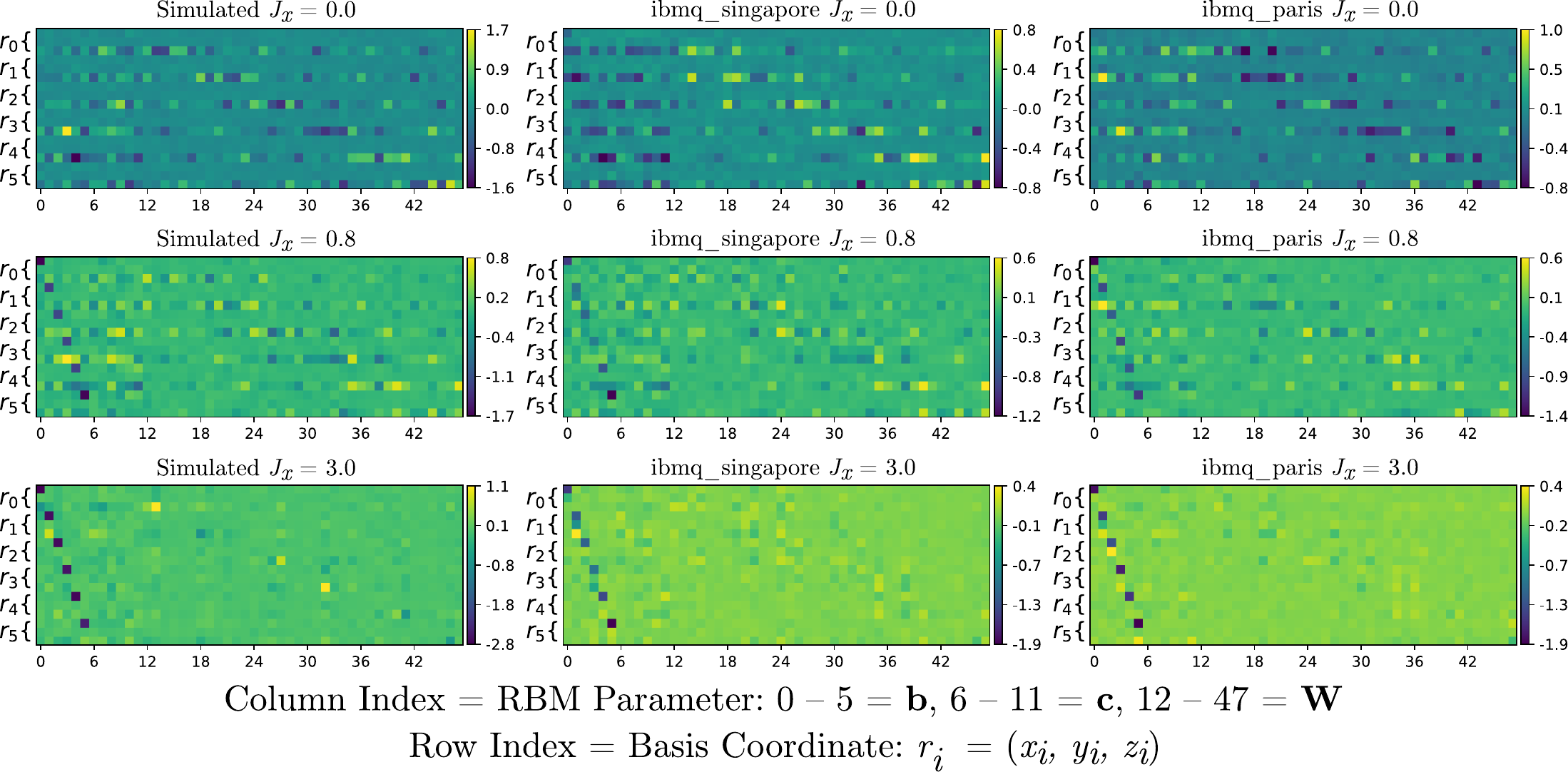}
\caption{\label{filter_pic}\textbf{Filters found for linear FFNNs trained on anti-ferromagnetic TFIM data.} These are visual representations of the component values of the FFNN filters found for some of the TFIM ground-states in section \ref{subsec:fid}. As a linear network was used, each component of a filter $M_{ij}$ gives the derivative of the predicted RBM parameter indexed $j$ (columns) with respect to a Bloch sphere coordinate $i$ (rows). Here the rows are grouped into coordinates $r_k=(x_k, y_k, z_k)$ to indicate which basis coordinates are associated with each qubit. A range of transverse fields are shown; $J_x=0$ being in the ordered, entangled phase, $J_x=3$ in the disordered low-entanglement phase, and $J_x=0.8$ which is in the intermediate phase around the phase transition. The RBM parameters (columns in each image) are indexed as follows; 0-5 are the biases on each visible neuron (site), 6-11 are the biases on each hidden neuron, and 12-47 are the (flattened) weights between visible and hidden neurons.}
\end{figure*}
The filters that were found for the simulated, $\textit{ibmq\_singapore}$, and $\textit{ibmq\_paris}$ data are shown in Figure \ref{filter_pic} for three values of $J_x$ (corresponding to the ordered, intermediate, and disordered phases). What is observed for the simulated data is broadly in line with predictions. For low transverse fields the majority of the filters’ non-zero components link the RBM parameters to the $z$ coordinates of the qubits’ measurement bases. In this regime significant weights between basis coordinates and hidden parameters (weights and biases) are present and the visible biases (the first 6 columns of each a filter) appear to have a delocalised dependence on the basis choice; these are all qualitative indicators of correlations, the latter delocalisation perhaps being an indicator of entanglement. As $J_x$ increases we see an increasingly prominent local dependency of the visible biases on the $x$ coordinates of each qubit’s measurement basis; in the intermediate regime both the non-local $z$ dependencies and local $x$ dependencies are present while for large transverse fields the filter components for the hidden weights and biases become suppressed and the dominant terms lie solely on the visible biases.\par 
Perhaps more interesting are the filters that were found for the two NISQ devices. The $\textit{ibmq\_singapore}$ device performed worse for large transverse fields than small ones. This is reflected in the filters as while the filter found for $J_x=0$ matches qualitatively well to that of the simulated data (with the same matrix elements being non-zero, although not necessarily having the exact same values), the filters for the larger fields begin to diverge quite dramatically. At large fields we still observe sizeable basis dependence of the hidden weights and biases (whereas these parameters should be close to zero), indicating that extra undesired correlations are present on the device. The visible biases in this region have a less clear basis dependence than for the simulated data; rather than strongly depending on $\{x_i\}$ and $\{z_i$\} (rows 0, 3, 6, ... and 2, 5, 8, ... respectively for each filter) the $y$ coordinates (rows 1, 4, 7, ...) of the bases also appreciably contribute. Unfortunately, as we are only looking at one family of states and the RBM parameters are hard to directly interpret it is difficult to conclusively pin down which processes are responsible for these discrepancies. The unexpected components corresponding to RBM weights (and so measurement correlations) could imply some kind of correlated noise process (for example cross-talk between qubits), whereas the smeared out visible bias basis dependence could indicate a locally acting error imposed by a qubits’ environment (a systematic single qubit rotation).\par 
As seen in section \ref{subsec:fid}, the $\textit{ibmq\_paris}$ device was a lot better at generating the desired ground-states. Little qualitative difference is seen between its filters and $\textit{ibmq\_singapore}$’s for low transverse fields (both agreeing reasonably well to that of the simulated data). Encouragingly, as the transverse field grows larger the $\textit{ibmq\_paris}$ device manages to do a decent job of correctly introducing the local $x$ dependencies of the visible biases and then damping down the correlations between qubits by reducing the basis dependence of the weights and hidden biases. As shown in Figure \ref{fig:fidelity_graphs}c, $\textit{ibmq\_paris}$ does not perfectly generate the desired states and this manifests itself as small differences in the filters. While it appears that the correct qualitative dependencies are present, the values of the components do differ and these differences are magnified by the non-linearities in the underlying RBM, leading to the predictive fidelity being around $5\%$ higher for the measured data than for the ideal case.\par
\section{Discussion}
In this work we have discussed a novel method by which machine-learning techniques may be used to perform quantum state sample tomography, a quantum state characterisation scheme that we posit is a far more tractable alternative to full QST. Characterisation of quantum states is an extremely useful tool in the analysis of NISQ devices however full tomography is prohibitively expensive in terms of the number of measurements required. The density matrix that full QST produces is also an unwieldy object for large systems due to its exponential size/number of parameters. Instead, we relax the requirement that a density matrix is produced and build a model for how the measurement distributions of the system vary with the choice of measurement basis, doing so in such a way that these distributions can be sampled from efficiently. The model is trained with measurements in random bases making as few assumptions about which measurements must be taken as possible. This allows it to perform with good predictive accuracy on NISQ devices where unpredictable errors and imperfect qubit control may move the state into unexpected regions of the Hilbert space. As no assumptions are made about the purity of the state (although this can be estimated through sampling \cite{Elben2020}) it is not restricted to pure states, furthering its applicability to NISQ devices.\par
Two types of neural network architecture are combined to yield this trainable model for the measurement distributions, the BDRBM. Provided the state possesses enough underlying structure, the model used to learn and represent thhe measurement distributions, a restricted Boltzmann machine, allows distributions for even very large systems to be expressed succinctly and sampled from efficiently. To account for the basis dependence of measurements that is integral to quantum mechanics the RBM model is controlled by a second trainable neural network (a FFNN) which assigns parameters to the RBM that depend on the input basis setting. ``Meta-learning" is an area of active interest, particularly in ML applications where only small datasets are feasibly available \cite{Finn2017}. This idea that one model's parameters can be training data for another model broadly fits into this field; however, it appears particularly well suited to the continuously varying measurement outcome distributions present in quantum mechanical systems.\par
We have also demonstrated how analysis of the learned parameters for the FFNN can be used to infer qualitative properties of the state. Comparisons between parameters learned for a simulated ideal state preparation and for those learned from a noisy device could be a useful tool in performing error analysis without having to perform prohibitively expensive process tomography. It also raises the question of whether it is possible to identify the Schmidt basis for a given state by iteratively relabelling the computational axis until the FFNN weights of the BDRBM representation of the state are maximally aligned along this direction, provided the state can be expressed using a sufficiently simple FFNN. By interpolating between the measurements used for training this tomographic scheme attempts to get as much information out of each measurement as possible allowing fewer measurements to be taken while still arriving at a useful generative model of the state's measurement statistics. This can then be either analysed directly, used to predict distributions in unseen bases, or efficiently draw samples in these bases. The predicted samples could then be used as an intermediate stage in optimisation schemes, allowing relevant quantities to be calculated without having to take additional measurements.\par
The scheme we have presented here has the potential complement existing quantum algorithms. For example, the ability to sample from the prepared state in arbitrary local bases would be of great use in the VQE algorithmn \cite{Kandala2017}, allowing estimates of the energy to be made while also providing extra information about the prepared state. Using the technique given in \cite{Elben2020} one can use samples in random local bases to estimate quantities of the form $\text{Tr}(\rho_1\rho_2)$ giving access to state fidelities of this form, the purity of the state, and its entanglement properties of the state (through Renyi entanglement entropies). Crucially, the ability to efficiently estimate the overlap between prepared states as well as their energy expectation values with only a modestly-sized set of random measurements could allow excited state generalisations of VQE \cite{Higgott2019} (based on projector methods) to be performed more efficiently than is currently feasible on-device -- such algorithms are highly relevant to quantum chemistry and materials simulations. This means that not only does our scheme provide a pathway to more efficient benchmarking and analysis of large device-prepared states but can also complement existing near-term algorithms.
\begin{acknowledgments}
This work has been supported by the UK EPSRC (EP/P510257/1), the EPSRC Hub in Quantum Computing and Simulation (EP/T001062/1), the Royal Society and the Samsung GRP grant. We acknowledge the use of IBM Quantum services for this work. The views expressed are those of the authors, and do not reflect the official policy or position of IBM or the IBM Quantum team. 
\end{acknowledgments}
\bibliography{references}
\end{document}